\documentclass[aps,pra,10pt,amsmath,twocolumn,letterpaper]{revtex4-1}
\usepackage{lmodern}
\usepackage{graphicx}

\usepackage[bookmarksdepth=2,colorlinks=true,linkcolor=blue,
            citecolor=red,filecolor=magenta,urlcolor=blue,
            breaklinks=true]{hyperref}

\begin{abstract}
This manuscript is a revision of our previous work that develops the three
layer model for the surface second-harmonic generation yield; here, we
add the necessary algebra to derive expressions that include elliptically
polarized incoming fields. This allows yet another degree of flexibility to the
previous work, as elliptical polarization is the most general polarization case
possible. The three layer model considers that the SH conversion takes place in
a thin layer just below the surface of a material. This layer lies under the
vacuum region, and above any bulk material that is not SHG active. The inherent
flexibility of this model makes it an excellent choice for thin films and 2D
materials.
\end{abstract}

\begin{document}

\title{General Theory for the Surface Second-Harmonic Generation Yield}
    \author{Sean M. Anderson}
    \email{sma@cio.mx}
    \affiliation{Centro de Investigaciones en \'Optica, A.C.,
                 Le\'on 37150, Mexico}
    \author{Yujin Cho}
    \affiliation{Department of Physics, University of Texas at Austin, Austin,
    Texas 78712, USA}

    \author{Bernardo S. Mendoza}
    \affiliation{Centro de Investigaciones en \'Optica, A.C.,
                 Le\'on 37150, Mexico}

\maketitle


\section{Introduction to the Three Layer Model}\label{sec:3layersshg}

This manuscript is a revision of our previous work featured in Refs.
\cite{andersonPRB16b} and \cite{andersonthesis}; here, we will derive the
formulas required for the calculation of the surface second-harmonic generation
(SSHG) yield including elliptically polarized incoming fields. This adds even
more flexibility to our framework, as we can now arbitrarily consider any
incoming polarization. The SSHG yield is defined as
\begin{equation}\label{eq:rintensities}
\mathcal{R}(\omega)=\frac{I(2\omega)}{I^2(\omega)},
\end{equation}
with the intensity (in MKS units \cite{boyd, sutherland}) given by
\begin{equation}\label{eq:intensity}
I(\omega)=
2\epsilon_{0}c\, n(\omega)|E(\omega)|^{2}
,
\end{equation}
where $n(\omega)=\sqrt{\epsilon(\omega)}$ is the index of refraction
($\epsilon(\omega)$ is the dielectric function), $\epsilon_{0}$ is the vacuum
permittivity, and $c$ the speed of light in vacuum.

\begin{figure}[b]
\centering 
\includegraphics[width=\linewidth]{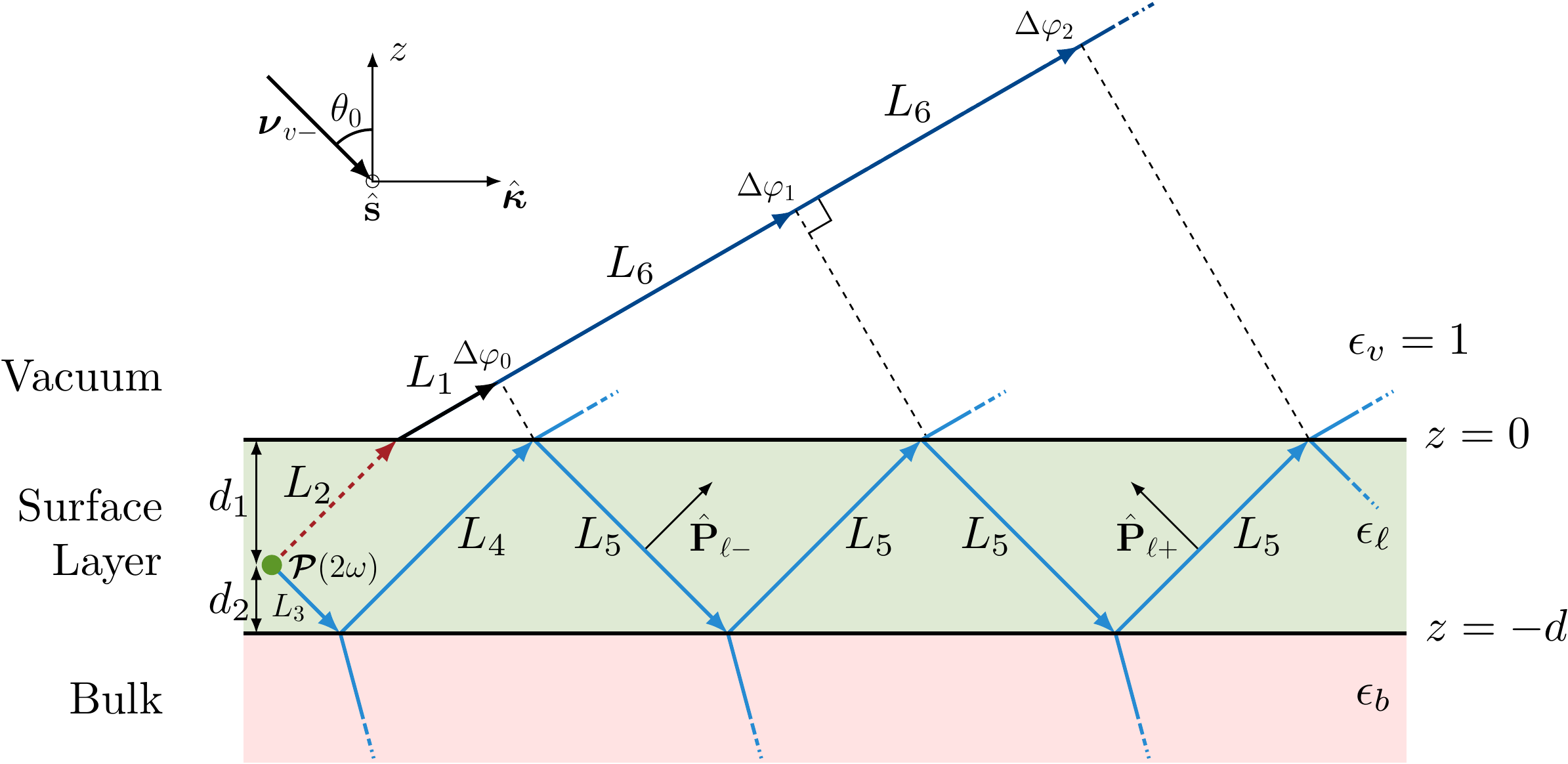}
\caption{Sketch of the three layer model for SHG.}
\label{fig:MR3layer2w}
\end{figure}

Our method for calculating $\mathcal{R}(\omega)$ is based on the work of Mizrahi
and Sipe \cite{mizrahiJOSA88}, since the derivation of the three layer model is
straightforward; see Fig. \ref{fig:MR3layer2w} for a detailed illustration of
this model. In this scheme, a given surface is represented by three regions or
layers. The first layer is the vacuum region (denoted by $v$) with a dielectric
function $\epsilon_{v}(\omega)=1$ from where the fundamental electric field
$\mathbf{E}_{v}(\omega)$ impinges on the material. The second layer is a
``thin'' layer (denoted by $\ell$) of thickness $d = d_{1} + d_{2}$
characterized by a dielectric function $\epsilon_{\ell}(\omega)$. It is in this
layer where the SH polarization sheet $\boldsymbol{\mathcal{P}}_{\ell}(2\omega)$
is located at $z_{\ell} = d_{1}$. The third layer is the bulk region denoted by
$b$ and characterized by $\epsilon_{b}(\omega)$. This bulk region can be made up
of any SHG inactive material (such as a substrate), which is why this model
readily lends itself to study thin films or 2D materials, as well as
conventional semiconductor surfaces. Both the vacuum and bulk layers are
semi-infinite.

The arrows in Fig. \ref{fig:MR3layer2w} point along the direction of
propagation, and the $p$-polarization unit vector, $\hat{\mathbf{P}}_{\ell
-(+)}$, along the downward (upward) direction is denoted with a thick arrow. The
$s$-polarization unit vector $\hat{\mathbf{s}}$, points out of the page. The
fundamental field $\mathbf{E}_{v}(\omega)$ is incident from the vacuum side
along the $\hat{\boldsymbol{\kappa}}z$-plane, with $\theta_{0}$ its angle of
incidence and $\boldsymbol{\nu}_{v-}$ its wave vector. $\Delta\varphi_{i}$
denotes the phase difference between the multiple reflected beams and the first
layer-vacuum transmitted beam, denoted by the dashed-red arrow (of length
$L_{2}$) followed by the solid black arrow (of length $L_{1}$). The dotted lines
in the vacuum region are perpendicular to the beam extended from the solid black
arrow (denoted by solid blue arrows of length $L_{6}$).

To begin our derivation of our model, we follow Ref. \cite{mizrahiJOSA88} and
assume a polarization sheet located at $z_{\beta}$, of the form
\begin{equation}\label{eq:psheet}
\mathbf{P}(\mathbf{r},t) = \boldsymbol{\mathcal{P}}
e^{i\boldsymbol{\kappa}\cdot\mathbf{R}}e^{-i\omega t}\delta(z - z_{\beta}) 
+ \mathrm{c.c.},
\end{equation}
where $\mathbf{R}=(x,y)$, $\boldsymbol{\kappa}$ is the component of the wave
vector $\boldsymbol{\nu}^{\phantom{a}}_{\beta}$ parallel to the surface, and
$z_{\beta}$ is the position of the sheet within medium $\beta$, and
$\boldsymbol{\mathcal{P}}$ is the position-independent polarization. Ref.
\cite{sipeJOSAB87} demonstrates that the solution of the Maxwell equations for
the radiated fields $E_{\beta,p\pm}$, and $E_{\beta,s}$ with
$\mathbf{P}(\mathbf{r},t)$ as a source at points $z\neq 0$, can be written as
\begin{equation}\label{eq:solmaxwell}
(E_{\beta,p\pm},E_{\beta,s}) = 
(\frac{\gamma i\tilde{\omega}^2}{2\epsilon_{0}\tilde{w}_{\beta}}
\,\hat{\mathbf{p}}_{\beta\pm}\cdot\boldsymbol{\mathcal{P}},
\frac{\gamma i\tilde{\omega}^2}{2\epsilon_{0}\tilde{w}_{\beta}}
\,\hat{\mathbf{s}}\cdot\boldsymbol{\mathcal{P}}),
\end{equation} 
where $\tilde{\omega}=\omega/c$; $\hat{\mathbf{s}}$ and
$\hat{\mathbf{p}}_{\beta\pm}$ are the unit vectors for the $s$ and $p$
polarizations of the radiated field, respectively. The $\pm$ refers to upward
($+$) or downward ($-$) direction of propagation within medium $\beta$, as
described in Fig. \ref{fig:MR3layer2w}. Also,
$\tilde{w}^{\phantom{a}}_{\beta}(\omega)=\tilde{\omega}w^{\phantom{a}}_{\beta}$,
where
\begin{equation}\label{eq:r4}
\hat{\mathbf{p}}^{\phantom{A}}_{\beta\pm}(\omega) =
  \frac{\kappa(\omega)\hat{\mathbf{z}}\mp 
  \tilde{w}^{\phantom{A}}_{\beta}(\omega)\hat{\boldsymbol{\kappa}}} 
  {\tilde{\omega} n^{\phantom{A}}_{\beta}(\omega)}
= \frac{\sin\theta_{0}\hat{\mathbf{z}}\mp 
  w^{\phantom{A}}_{\beta}(\omega)\hat{\boldsymbol{\kappa}}} 
  {n^{\phantom{A}}_{\beta}(\omega)},
\end{equation}
with
\begin{equation}\label{eq:wavevector}
w^{\phantom{a}}_{\beta}(\omega) = 
\big(\epsilon^{\phantom{a}}_{\beta}(\omega) - \sin^{2}\theta_{0}\big)^{1/2},
\end{equation}
$\theta_{0}$ is the angle of incidence of $\mathbf{E}_{v}(\omega)$,
$\kappa(\omega)=\vert\boldsymbol{\kappa}\vert = \tilde{\omega}\sin\theta_{0}$,
$n^{\phantom{A}}_{\beta}(\omega)=\sqrt{\epsilon^{\phantom{A}}_{\beta}(\omega)}$
is the index of refraction of medium $\beta$, and $z$ is the direction
perpendicular to the surface that points towards the vacuum. If we consider the
plane of incidence along the $\boldsymbol{\kappa}z$ plane, then
\begin{equation}\label{eq:mc1}
\hat{\boldsymbol{\kappa}} = \cos\phi\hat{\mathbf{x}} + \sin\phi\hat{\mathbf{y}},
\end{equation}
and
\begin{equation}\label{eq:mmc2}
\hat{\mathbf{s}} = -\sin\phi\hat{\mathbf{x}} + \cos\phi\hat{\mathbf{y}},
\end{equation}
where $\phi$ is the azimuthal angle with respect to the $x$ axis.

The nonlinear polarization responsible for the SHG is immersed in the thin layer
($\beta=\ell$), and is given by
\begin{equation}\label{eq:tres}
\mathcal{P}^{\mathrm{a}}_{\ell}(2\omega) =
\epsilon_{0}\chi^{\mathrm{abc}}_{\mathrm{surface}}(-2\omega;\omega,\omega)
    E^{\mathrm{b}}(\omega)E^{\mathrm{c}}(\omega)
,
\end{equation}
where $\boldsymbol{\chi}_{\mathrm{surface}}(-2\omega;\omega,\omega)$ is the
dipolar surface nonlinear susceptibility tensor; the calculation of this
quantity is given in detail in Refs. \cite{andersonthesis} and
\cite{andersonPRB15}. We will omit the $(-2\omega;\omega,\omega)$ notation from
this point on. The Cartesian indices $\mathrm{a,b,c}$ are summed over if
repeated; $\chi^{\mathrm{abc}} = \chi^{\mathrm{acb}}$ is the intrinsic
permutation symmetry due to the fact that SHG is degenerate in
$E^{\mathrm{b}}(\omega)$ and $E^{\mathrm{c}}(\omega)$. As in Ref.
\cite{mizrahiJOSA88}, we consider the polarization sheet (Eq. \eqref{eq:psheet})
to be oscillating at some frequency $\omega$ in order to properly express Eqs.
\eqref{eq:solmaxwell}--\eqref{eq:mmc2}. However, in the following we find it
convenient to use $\omega$ exclusively to denote the fundamental frequency and
$\boldsymbol{\kappa}$ to denote the component of the incident wave vector
parallel to the surface. The generated nonlinear polarization is oscillating at
$\Omega = 2\omega$ and will be characterized by a wave vector parallel to the
surface $\mathbf{K} = 2\boldsymbol{\kappa}$. We can carry over Eqs.
\eqref{eq:psheet}--\eqref{eq:mmc2} simply by replacing the lowercase symbols
($\omega,\tilde{\omega},\boldsymbol{\kappa},n^{\phantom{A}}_{\beta},
\tilde{w}^{\phantom{A}}_{\beta},w^{\phantom{A}}_{\beta},
\hat{\mathbf{p}}^{\phantom{A}}_{\beta\pm},\hat{\mathbf{s}}$) with uppercase
symbols ($\Omega,\tilde{\Omega},\mathbf{K},N^{\phantom{A}}_{\beta},
\tilde{W}^{\phantom{A}}_{\beta},W^{\phantom{A}}_{\beta},
\hat{\mathbf{P}}_{\beta\pm},\hat{\mathbf{S}}$), all evaluated at $2\omega$. Of
course, we always have that $\hat{\mathbf{S}}=\hat{\mathbf{s}}$.

From Fig. \ref{fig:MR3layer2w}, we observe the propagation of the SH field as it
is refracted at the layer-vacuum interface ($\ell v$), and  reflected multiple
times from the layer-bulk ($\ell b$) and layer-vacuum ($\ell v$) interfaces.
Thus, we can define
\begin{equation}\label{eq:r5}
\mathbf{T}^{\ell v}
= \hat{\mathbf{s}}T_{s}^{\ell v}\hat{\mathbf{s}} 
+ \hat{\mathbf{P}}_{v+}T_{p}^{\ell v} \hat{\mathbf{P}}_{\ell +},
\end{equation}
as the transmission tensor for the $\ell v$ interface,
\begin{equation}\label{eq:r6}
\mathbf{R}^{\ell b}
= \hat{\mathbf{s}}R_{s}^{\ell b}\hat{\mathbf{s}}
+ \hat{\mathbf{P}}_{\ell +}R_{p}^{\ell b} \hat{\mathbf{P}}_{\ell -},
\end{equation} 
as the reflection tensor for the $\ell b$ interface, and
\begin{equation}\label{eq:r6b}
\mathbf{R}^{\ell v}
= \hat{\mathbf{s}}R_{s}^{\ell v}\hat{\mathbf{s}}
+ \hat{\mathbf{P}}_{\ell -}R_{p}^{\ell v} \hat{\mathbf{P}}_{\ell +},
\end{equation} 
as the reflection tensor for the $\ell v$ interface. The Fresnel factors in
uppercase letters, $T^{ij}_{s,p}$ and $R^{ij}_{s,p}$, are evaluated at $2\omega$
from the following well known formulas \cite{jacksonbook},
\begin{equation*}\label{eq:e.f1}
\begin{split}
t_{s}^{ij}(\omega) &=
\frac{2w_{i}(\omega)}{w_{i}(\omega) + w_{j}(\omega)},\\
t_{p}^{ij}(\omega) &=
\frac{2w_{i}(\omega)\sqrt{\epsilon_{i}(\omega)\epsilon_j(\omega)}}
     {w_{i}(\omega)\epsilon_{j}(\omega) + w_{j}(\omega)\epsilon_{i}(\omega)},\\
r_{s}^{ij}(\omega) &=
\frac{w_{i}(\omega) - w_{j}(\omega)}
     {w_{i}(\omega) + w_{j}(\omega)},\\
r_{p}^{ij}(\omega) &=
\frac{w_{i}(\omega)\epsilon_{j}(\omega) - w_{j}\epsilon_{i}(\omega)}
     {w_{i}(\omega)\epsilon_{j}(\omega) + w_{j}(\omega)\epsilon_{i}(\omega)}. 
\end{split}
\end{equation*}
With these expressions we easily derive the following useful relations,
\begin{equation}\label{eq:mf}
\begin{split}
1 + r^{\ell b}_{s} &= t^{\ell b}_{s},\\
1 + r^{\ell b}_{p} &= \frac{n_{b}}{n_{\ell}}t^{\ell b}_{p},\\
1 - r^{\ell b}_{p} &= \frac{n_{\ell}}{n_{b}}\frac{w_{b}}{w_{\ell}}
                      t^{\ell b}_{p},\\
t^{\ell v}_{p} &= \frac{w_{\ell}}{w_{v}}t^{v\ell}_{p},\\
t^{\ell v}_{s} &= \frac{w_{\ell}}{w_{v}}t^{v\ell}_{s}.
\end{split}
\end{equation}


\section{Multiple Reflections}


\subsection{Multiple SHG Reflections}

The SH field $\mathbf{E}(2\omega)$ radiated by the SH polarization
$\boldsymbol{\mathcal{P}}_{\ell}(2\omega)$ will radiate directly into the vacuum
and the bulk, where it will be reflected back at the layer-bulk interface into
the thin layer. This beam will be transmitted and reflected multiple times, as
shown in Fig. \ref{fig:MR3layer2w}. As the two beams propagate, a phase
difference will develop between them according to
\begin{equation*}\label{eq:m99}
\begin{split}
\Delta\varphi_{m} 
&= \tilde{\Omega}
\Big(
(L_{3} + L_{4} + 2mL_{5})N_{\ell}\\
&\qquad\qquad- \big(L_{2}N_{\ell} + (L_{1} + mL_{6})N_{v}\big)
\Big)\\
&= \delta_{0} + m\delta,\quad m=0,1,2,\ldots,
\end{split}
\end{equation*}
where
\begin{equation}\label{eq:delta0}
\delta_{0} =
8\pi\left(\frac{d_{2}}{\lambda_{0}}\right)W_{\ell},
\end{equation}
and
\begin{equation}\label{eq:delta}
\delta = 8\pi
\left(\frac{d}{\lambda_{0}}\right)W_{\ell},
\end{equation}
where $\lambda_{0}$ is the wavelength of the fundamental field in the vacuum,
$W_{\ell}$ is described in Eq. \eqref{eq:wavevector}, $d$ is the thickness of
layer $\ell$, and $d_{2}$ is the distance between
$\boldsymbol{\mathcal{P}}_{\ell}(2\omega)$ and the $\ell b$ interface (see Fig.
\ref{fig:MR3layer2w}). We see that $\delta_{0}$ is the phase difference of the
first and second transmitted beams, and $m\delta$ that of the first and third
($m = 1$), first and fourth ($m = 2$), and so on. Note that the thickness $d$ of
the layer $\ell$ enters through the phase $\delta$, and the position $d_{2}$ of
the nonlinear polarization $\mathbf{P}(\mathbf{r},t)$ (Eq. \eqref{eq:psheet})
enters through $\delta_{0}$. In particular, $d_{2}$ could be used as a variable
to study the effects of multiple reflections on the SSHG yield
$\mathcal{R}(2\omega)$.

To take into account the multiple reflections of the generated SH field in the
layer $\ell$, we proceed as follows. We only include the algebra for the
$p$-polarized SH field, though the $s$-polarized field could be worked out along
the same steps. The $p$-polarized $\mathbf{E}_{\ell,p}(2\omega)$ field reflected
multiple times is given by
\begin{widetext}
\begin{equation*}\label{eq:E2wcomplete}
\begin{split}
\mathbf{E}_{\ell,p}(2\omega) 
&= E_{\ell,p+}(2\omega)\mathbf{T}^{\ell v}\cdot\hat{\mathbf{P}}_{\ell +}
 + E_{\ell,p-}(2\omega)\mathbf{T}^{\ell v}
\cdot\mathbf{R}^{\ell b}\cdot\hat{\mathbf{P}}_{\ell-}e^{i\Delta\varphi_{0}}\\
&+ E_{\ell,p-}(2\omega)\mathbf{T}^{\ell v}
\cdot\mathbf{R}^{\ell b}\cdot\mathbf{R}^{\ell v}
\cdot\mathbf{R}^{\ell b}\cdot\hat{\mathbf{P}}_{\ell-}e^{i\Delta\varphi_{1}}
\\
&+ E_{\ell,p-}(2\omega)\mathbf{T}^{\ell v}
\cdot\mathbf{R}^{\ell b}\cdot\mathbf{R}^{\ell v}
\cdot\mathbf{R}^{\ell b}\cdot\mathbf{R}^{\ell v}
\cdot\mathbf{R}^{\ell b}\cdot\hat{\mathbf{P}}_{\ell-}e^{i\Delta\varphi_{2}}
+\cdots\\
&= E_{\ell,p+}(2\omega)\mathbf{T}^{\ell v}\cdot\hat{\mathbf{P}}_{\ell +}
+ E_{\ell,p-}(2\omega) \mathbf{T}^{\ell v}
\cdot\sum_{m=0}^\infty  
\big(
\mathbf{R}^{\ell b}\cdot\mathbf{R}^{\ell v} 
e^{i\delta}\Big)^m 
\cdot\mathbf{R}^{\ell b}\cdot\hat{\mathbf{P}}_{\ell-}e^{i\delta_{0}}.
\end{split}
\end{equation*}
\end{widetext}
From Eqs. \eqref{eq:r5}--\eqref{eq:r6b} it is easy to show that
\begin{equation*}\label{eq:m1}
\begin{split}
\mathbf{T}^{\ell v}\cdot
\Big(\mathbf{R}^{\ell b}\cdot\mathbf{R}^{\ell v}\Big)^{n}\cdot
\mathbf{R}^{\ell b}
&= \hat{\mathbf{s}}T^{\ell v}_{s}
  \Big(R^{\ell b}_{s}R^{\ell v}_{s}\Big)^{n}R^{\ell b}_{s}\hat{\mathbf{s}}\\
&+ \hat{\mathbf{P}}_{v+}T^{\ell v}_{p}\Big(R^{\ell b}_{p}R^{\ell v}_{p}\Big)^n 
  R^{\ell b}_{p} 
\hat{\mathbf{P}}_{\ell-},
\end{split}
\end{equation*}
then,
\begin{equation*}\label{eq:E2wreduced}
\begin{split}
\mathbf{E}_{\ell,p}(2\omega) 
&= \hat{\mathbf{P}}_{\ell +}T^{\ell v}_{p}
\Big(
E_{\ell,p+}(2\omega)\\
&\hspace{2cm}+
\frac{R^{\ell b}_{p}e^{i\delta_{0}}}{1 + R^{v\ell}_{p}R^{\ell b}_{p}e^{i\delta}}
E_{\ell,p-}(2\omega) 
\Big),
\end{split}
\end{equation*}
where we used $R^{ij}_{s,p} = -R^{ji}_{s,p}$. Using Eq. \eqref{eq:solmaxwell}
and \eqref{eq:mf}, we can readily write
\begin{equation}\label{eq:mr8}
\mathbf{E}_{\ell,p}(2\omega) =
\frac{\gamma i\tilde{\Omega}}{W_{\ell}}\mathbf{H}_{\ell}\cdot
\boldsymbol{\mathcal{P}}_{\ell}(2\omega),
\end{equation}
where
\begin{equation}\label{eq:mr9}
\begin{split}
\mathbf{H}_{\ell}
&= \frac{W_\ell}{W_v}
\Bigg[
\hat{\mathbf{s}}\,T_{s}^{v\ell}
\left(1+ R^{M}_{s}\right)\hat{\mathbf{s}}\\
&\hspace{1.5cm}+ \hat{\mathbf{P}}_{v+}T_{p}^{v\ell}
\left(\hat{\mathbf{P}}_{\ell +} + R^{M}_{p}\hat{\mathbf{P}}_{\ell -}\right)
\Bigg],
\end{split}
\end{equation}
and
\begin{equation}\label{m61}
R^{M}_{\mathrm{i}}\equiv
\frac{R^{\ell b}_{\mathrm{i}}e^{i\delta_{0}}}
     {1+R^{v\ell}_{\mathrm{i}} R^{\ell b}_{\mathrm{i}}e^{i\delta}},
     \quad \mathrm{i}=s,p,
\end{equation}
is defined as the multiple ($M$) reflection coefficient. This coefficient
depends on the thickness $d$ of layer $\ell$, and most importantly on the
position $d_{2}$ of $\boldsymbol{\mathcal{P}}_{\ell}(2\omega)$ within this
layer. The final results will depend on both $d$ and $d_{2}$. However, using Eq.
\eqref{eq:delta0} we can also define an average $\bar{R}^{M}_{\mathrm{i}}$ as
\begin{equation*}\label{eq:mcave}
\begin{split}
\bar{R}^{M}_{\mathrm{i}}
&\equiv
\frac{1}{d}\int_{0}^{d}
\frac{R^{\ell b}_{\mathrm{i}}e^{i(8\pi W_{\ell}/\lambda_{0})x}}
{1 + R^{v\ell}_{\mathrm{i}}R^{\ell b}_{\mathrm{i}}e^{i\delta}}\,dx\\\\
&= \frac{R^{\ell b}_{\mathrm{i}}e^{i\delta/2}}
{1 + R^{v\ell}_{\mathrm{i}}R^{\ell b}_{\mathrm{i}}e^{i\delta}}
\,\mathrm{sinc}(\delta/2),
\end{split}
\end{equation*}
that only depends on $d$ through the $\delta$ term from Eq. \eqref{eq:delta}. It
is very convenient to go ahead and define
\begin{equation}\label{eq:rm}
R^{M\pm}_{\mathrm{i}}\equiv 1 \pm R^{M}_{\mathrm{i}}, \quad \mathrm{i}=s,p.
\end{equation}

To connect with the work in Ref. \cite{mizrahiJOSA88}, where
$\boldsymbol{\mathcal{P}}(2\omega)$ is located on top of the vacuum-surface
interface and only the vacuum radiated beam and the first (and only) reflected
beam need be considered, we take $\ell = v$ and $d_{2} = 0$, then $T^{\ell v} =
1$, $R^{v\ell} = 0$ and $\delta_{0} = 0$, with which $R^{M}_{\mathrm{i}} =
R^{vb}_{\mathrm{i}}$. Thus, Eq. \eqref{eq:mr9} coincides with Eq. (3.8) of Ref.
\cite{mizrahiJOSA88}.


\subsection{Multiple Reflections for the Linear Field}

\begin{figure}[t]
\centering 
\includegraphics[width=\linewidth]{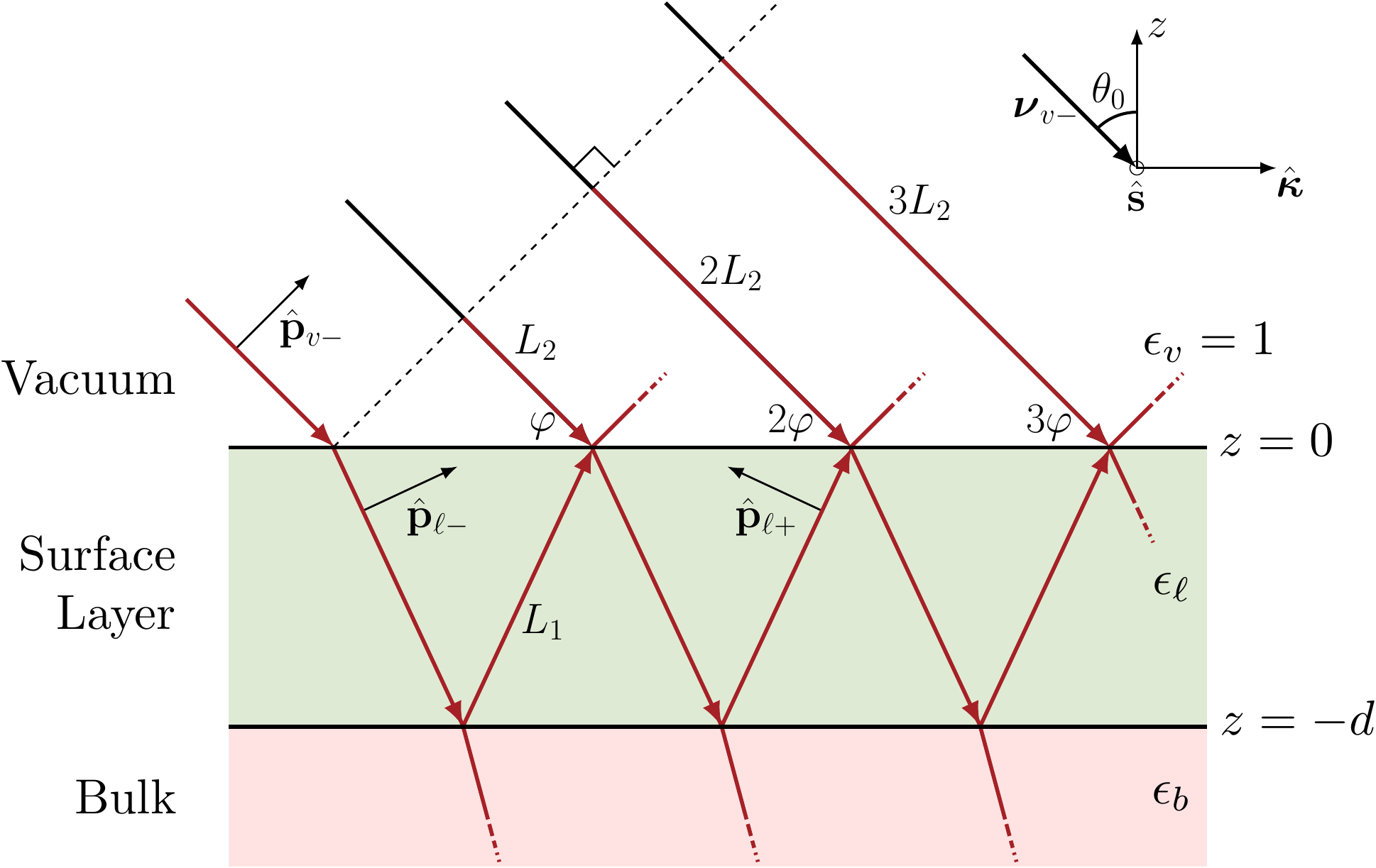}
\caption{Sketch of the multiple reflected fundamental fields in the three layer
model.}\label{fig:MR3layer1w}
\end{figure}

We must also consider the multiple reflections of the fundamental field
$\mathbf{E}_{\ell}(\omega)$ inside the thin layer $\ell$. In Fig.
\ref{fig:MR3layer1w} we present the situation where $\mathbf{E}_{v}(\omega)$
impinges from the vacuum side along the $\hat{\boldsymbol{\kappa}}z$-plane.
$\theta_{0}$ and $\boldsymbol{\nu}_{v-}$ are the angle of incidence and wave
vector, respectively. The arrows point along the direction of propagation. The
$p$-polarization unit vectors $\hat{\mathbf{p}}_{\beta\pm}$, point along the
downward $(-)$ or upward $(+)$ directions and are denoted with thick arrows,
where $\beta = v$ or $\ell$. The $s$-polarization unit vector $\hat{\mathbf{s}}$
points out of the page.

As the first transmitted beam is reflected multiple times from the $\ell b$ and
the $\ell v$ interfaces, it accumulates a phase difference of $n\varphi$ (with
$n=1,2,3,\ldots$) with respect to the incident field. $\varphi$ is given by
\begin{equation}\label{mphi}
\begin{split}
\varphi &= \frac{\omega}{c}(2L_{1}n_{\ell} - L_{2}n_{v}) 
= 4\pi\left(\frac{d}{\lambda_{0}}\right)w_{\ell},
\end{split}
\end{equation}
where $n_{v} = 1$. We need Eqs. \eqref{eq:r6} and \eqref{eq:r6b} for $1\omega$,
and also need
\begin{equation*}\label{eq:mvv}
\mathbf{t}^{v\ell}
= \hat{\mathbf{s}}t_{s}^{v\ell}\hat{\mathbf{s}} 
+ \hat{\mathbf{p}}_{\ell -}t_{p}^{v\ell}\hat{\mathbf{p}}_{v-},
\end{equation*}
to write
\begin{widetext}
\begin{equation*}\label{eq:mcvew}
\begin{split}
\mathbf{E}_{\ell}(\omega)
&= E_{0}
\Big[
\mathbf{t}^{v\ell} + \mathbf{r}^{\ell b}\cdot\mathbf{t}^{v\ell}e^{i\varphi}
 + \mathbf{r}^{\ell b}\cdot\mathbf{r}^{\ell v}\cdot
   \mathbf{r}^{\ell b}\cdot\mathbf{t}^{v\ell} e^{i2\varphi}
 + \mathbf{r}^{\ell b}\cdot\mathbf{r}^{\ell v}\cdot
   \mathbf{r}^{\ell b}\cdot\mathbf{r}^{\ell v}\cdot
   \mathbf{r}^{\ell b}\cdot\mathbf{t}^{v\ell} e^{i3\varphi}
 + \cdots
\Big]\cdot\hat{\mathbf{e}}^{\mathrm{i}}\nonumber\\
&= E_{0}
\Big[
1 + \Big(1 + \mathbf{r}^{\ell b}\cdot\mathbf{r}^{\ell v}e^{i\varphi}
+ (\mathbf{r}^{\ell b}\cdot\mathbf{r}^{\ell v})^2e^{i2\varphi}+\cdots\Big)\cdot
\mathbf{r}^{\ell b}e^{i\varphi}
\Big]
\cdot\mathbf{t}^{v\ell}\cdot\hat{\mathbf{e}}^{\mathrm{i}}\nonumber\\
&= E_{0}
\Big[
\hat{\mathbf{s}} t^{v\ell}_{s}(1+r^{M}_{s})\hat{\mathbf{s}} 
+ t^{v\ell}_{p}
\left(\hat{\mathbf{p}}_{\ell-}+\hat{\mathbf{p}}_{\ell+}r^{M}_{p}\right)
\hat{\mathbf{p}}_{v-}
\Big]\cdot\hat{\mathbf{e}}^{\mathrm{i}},
\end{split}
\end{equation*}
where $E_{0}$ is the intensity of the fundamental field, and
$\hat{\mathbf{e}}^{\mathrm{i}}$ is the unit vector of the incoming polarization
($\mathrm{i} = s,p$), with $\hat{\mathbf{e}}^{s}=\hat{\mathbf{s}}$ and
$\hat{\mathbf{e}}^{p}=\hat{\mathbf{p}}_{v-}$. Also,
\begin{equation}\label{mvrm}
r^{M}_{\mathrm{i}} \equiv
\frac{r^{\ell b}_{\mathrm{i}}e^{i\varphi}}{1+r^{v\ell}_{\mathrm{i}}
r^{\ell b}_{\mathrm{i}}e^{i\varphi}}, \quad \mathrm{i}=s,p.
\end{equation}
$r^{M}_{\mathrm{i}}$ is defined as the multiple (M) reflection coefficient for
the fundamental field. We define $\mathbf{E}^{\mathrm{i}}_{\ell}(\omega)\equiv
E_{0}\mathbf{e}^{\omega,\mathrm{i}}_{\ell}$ ($\mathrm{i}=s,p$), where
\begin{equation*}\label{eq:mcvew2}
\mathbf{e}^{\omega,\mathrm{i}}_\ell 
= \Big[\hat{\mathbf{s}} t^{v\ell}_s(1+r^M_s)\hat{\mathbf{s}} 
+ t^{v\ell}_p\left(\hat{\mathbf{p}}_{\ell-}+\hat{\mathbf{p}}_{\ell+}r^{M}_p 
\right)\hat{\mathbf{p}}_{v-}
\Big]\cdot\hat{\mathbf{e}}^{\mathrm{i}},
\end{equation*}
and using Eqs. \eqref{eq:r4}, \eqref{eq:mc1}, and \eqref{eq:mmc2} we obtain that
\begin{equation}\label{eq:vec1wcomplete}
\mathbf{e}^{\omega,i}_{\ell}
=
\Bigg[
t^{v\ell}_{s} r^{M+}_{s}
\left(- \sin\phi\hat{\mathbf{x}} + \cos\phi\hat{\mathbf{y}}\right)
\hat{\mathbf{s}}
+
\frac{t^{v\ell}_{p}}{n_{\ell}}
\left( 
  r^{M+}_{p}\sin\theta_{0}\hat{\mathbf{z}}
+ r^{M-}_{p}w_{\ell}\cos\phi\hat{\mathbf{x}}
+ r^{M-}_{p}w_{\ell}\sin\phi\hat{\mathbf{y}}
\right)
\hat{\mathbf{p}}_{v-}
\Bigg]
\cdot \hat{\mathbf{e}}^{\mathrm{i}},
\end{equation} 
\end{widetext}
where
\begin{equation*}\label{eq:mvc}
r^{M\pm}_{\mathrm{i}}=1\pm r^{M}_{\mathrm{i}},\quad \mathrm{i} = s,p.
\end{equation*}

\section{General Polarization Considerations}


\subsection{\texorpdfstring{$2\omega$}{2w} Terms for \emph{P} and \emph{S}
Linear Polarization}

The outgoing SHG radiation will almost always be measured in some configuration
of $P$ and $S$ polarization.
Using Eq. \eqref{eq:mf}, we can write Eq. \eqref{eq:mr8} as
\begin{equation}\label{eq:r10}
\begin{split}
E_{\ell}(2\omega)
&= \frac{2\gamma i\omega}{cW_\ell}
\hat{\mathbf{e}}^{\mathrm{F}}\cdot\mathbf{H}_{\ell}\cdot
\boldsymbol{\mathcal{P}}_{\ell}(2\omega)\\\\
&= \frac{2\gamma i\omega}{cW_{v}}
\mathbf{e}^{\,2\omega,\mathrm{F}}_{\ell}\cdot
\boldsymbol{\mathcal{P}}_{\ell}(2\omega),
\end{split}
\end{equation}
where
\begin{equation}\label{eq:r12mm}
\begin{split}
\mathbf{e}^{2\omega,\mathrm{F}}_{\ell}
&= \hat{\mathbf{e}}^{\mathrm{F}}\cdot 
\Bigg[
\hat{\mathbf{s}}T_{s}^{v\ell}R^{M+}_{s}\hat{\mathbf{s}}\\
&+ \hat{\mathbf{P}}_{v+}
\frac{T^{v\ell}_{p}}
     {N_{\ell}}
\left(
\sin\theta_{0}R^{M+}_{p}\hat{\mathbf{z}}
- W_{\ell}R^{M-}_{p}\hat{\boldsymbol{\kappa}}
\right)
\Bigg],
\end{split}
\end{equation}
remembering that $R^{M\pm}_{p}$ was previously defined in Eq. \eqref{eq:rm}. By
substituting Eqs. \eqref{eq:mc1} and \eqref{eq:mmc2} into Eq.
\eqref{eq:r12mm}, we obtain
\begin{equation}\label{eq:e2wpmr}
\begin{split}
\mathbf{e}^{2\omega,P}_{\ell} =
\frac{T^{v\ell}_{p}}{N_{\ell}}
\big(
&- W_{\ell}R^{M-}_{p}\cos\phi\hat{\mathbf{x}}\\
&- W_{\ell}R^{M-}_{p}\sin\phi\hat{\mathbf{y}}
  \sin\theta_{0}R^{M+}_{p}\hat{\mathbf{z}}
\big),
\end{split}
\end{equation}
for $P$ $(\hat{\mathbf{e}}^{\mathrm{F}} = \hat{\mathbf{P}}_{v+})$ outgoing
polarization, and
\begin{equation}\label{eq:e2wsmr}
\mathbf{e}^{2\omega,S}_{\ell} =
T_{s}^{v\ell}R^{M+}_{s}
\left(
- \sin\phi\hat{\mathbf{x}}
+ \cos\phi\hat{\mathbf{y}}
\right).
\end{equation}
for $S$ $(\hat{\mathbf{e}}^{\mathrm{F}} = \hat{\mathbf{s}})$ outgoing
polarization.


\subsection{\texorpdfstring{$1\omega$}{1w} Terms for for Elliptical
Polarization}

Up until this juncture, we have not assumed any given polarization for the
incoming fields, other than that they must be in some combination of $p$ or $s$
polarization. But let us consider the most general polarization case, elliptical
polarization, by establishing that \cite{byersPRB94}
\begin{equation}\label{eq:generalpol}
\hat{\mathbf{e}}^{\mathrm{i}}
= \sin\alpha\,\hat{\mathbf{s}}
+ e^{i\tau}\cos\alpha\,\hat{\mathbf{p}}_{v-}.
\end{equation}
\begin{widetext}
Plugging this into Eq. \eqref{eq:vec1wcomplete} yields
\begin{equation}\label{eq:newpol}
\mathbf{e}^{\omega}_{\ell}
=
\Bigg[
\sin\alpha\,
t^{v\ell}_{s} r^{M+}_{s}
\left(- \sin\phi\hat{\mathbf{x}} + \cos\phi\hat{\mathbf{y}}\right)
+
e^{i\tau}\cos\alpha\,
\frac{t^{v\ell}_{p}}{n_{\ell}}
\left( 
  r^{M+}_{p}\sin\theta_{0}\hat{\mathbf{z}}
+ r^{M-}_{p}w_{\ell}\cos\phi\hat{\mathbf{x}}
+ r^{M-}_{p}w_{\ell}\sin\phi\hat{\mathbf{y}}
\right)
\Bigg].
\end{equation}
Thinking ahead, it will be very hand to have the expression for
$\mathbf{e}^{\omega}_{\ell}\mathbf{e}^{\omega}_{\ell}$. Multiplying these terms
out leads to the following expression,
\begin{equation}\label{eq:nonmatrix}
\begin{split}
\mathbf{e}^{\omega}_{\ell}
\mathbf{e}^{\omega}_{\ell}
&=
\sin^{2}\alpha\,
\left(t^{v\ell}_{s}r^{M+}_{s}\right)^{2}
\big(
  \sin^{2}\phi\hat{\mathbf{x}}\hat{\mathbf{x}}
 + \cos^{2}\phi\hat{\mathbf{y}}\hat{\mathbf{y}}
 - 2\sin\phi\cos\phi\hat{\mathbf{x}}\hat{\mathbf{y}}
\big)\\
&+
e^{2i\tau} \cos^{2}\alpha\,
\left(\frac{t^{v\ell}_{p}}{n_{\ell}}\right)^{2}
\bigg(
\big(r^{M-}_{p}\big)^{2}w^{2}_{\ell}\cos^{2}\phi
    \hat{\mathbf{x}}\hat{\mathbf{x}}
  + \big(r^{M-}_{p}\big)^{2}w^{2}_{\ell}\sin^{2}\phi
    \hat{\mathbf{y}}\hat{\mathbf{y}}
  + \big(r^{M+}_{p}\big)^{2}\sin^{2}\theta_{0}
    \hat{\mathbf{z}}\hat{\mathbf{z}}\\
  &\hspace{3.5cm}+ 2r^{M+}_{p}r^{M-}_{p}w_{\ell}\sin\theta_{0}\sin\phi
    \hat{\mathbf{y}}\hat{\mathbf{z}}
  + 2r^{M+}_{p}r^{M-}_{p}w_{\ell}\sin\theta_{0}\cos\phi
    \hat{\mathbf{x}}\hat{\mathbf{z}}
  + 2\big(r^{M-}_{p}\big)^{2}w^{2}_{\ell}\sin\phi\cos\phi
    \hat{\mathbf{x}}\hat{\mathbf{y}}
\bigg)\\
&+
2e^{i\tau}\sin\alpha\cos\alpha\,
\frac{t^{v\ell}_{p}t^{v\ell}_{s}r^{M+}_{s}}{n_{\ell}}
\Big(
  - r^{M-}_{p}w_{\ell}\sin\phi\cos\phi
    \hat{\mathbf{x}}\hat{\mathbf{x}}
  + r^{M-}_{p}w_{\ell}\sin\phi\cos\phi
    \hat{\mathbf{y}}\hat{\mathbf{y}}\\
  &\hspace{4.5cm}+ r^{M+}_{p}\sin\theta_{0}\cos\phi
    \hat{\mathbf{y}}\hat{\mathbf{z}}
  - r^{M+}_{p}\sin\theta_{0}\sin\phi
    \hat{\mathbf{x}}\hat{\mathbf{z}}
  + r^{M-}_{p}w_{\ell}\cos 2\phi
    \hat{\mathbf{x}}\hat{\mathbf{y}}
\Big).
\end{split}
\end{equation}
\end{widetext}

Given that the terms for $1\omega$ are now presented for the most general
polarization case, we can easily recover the expressions for $p$ and $s$ linear
polarization by plugging in the appropriate values for $\alpha$ and $\tau$
(featured in Table \ref{tab:polcases}) into Eq. \eqref{eq:nonmatrix}.

\begin{table}[t]
\caption{Values for $\alpha$ and $\tau$ (see Eq. \eqref{eq:generalpol}) that
yield common polarization cases.\label{tab:polcases}}
\begin{tabular}{ | l l | c c | }
\hline
\multicolumn{2}{|c|}{Type}                    & $\alpha$  & $\tau$    \\
\hline
Linear        & $p$ ($\hat{\mathbf{p}}_{v-}$) & 0         & 0         \\
Linear        & $s$ ($\hat{\mathbf{s}}$)      & $\pi/2$   & 0         \\
Linear        & $p$ + $s$ & $\pi/4$           & 0         \\
\hline
Circular      & Left                          & $\pi/4$   & $-\pi/2$  \\
Circular      & Right                         & $\pi/4$   & $+\pi/2$  \\
\hline
Elliptical    &                               & Any       & Any       \\
\hline
\end{tabular}
\end{table}


\section{The SSHG Yield}

The magnitude of the radiated field is given by $E(2\omega) =
\hat{\mathbf{e}}^{\mathrm{F}}\cdot\mathbf{E}_{\ell}(2\omega)$, where
$\hat{\mathbf{e}}^{\mathrm{F}}$ is the unit vector of the final SH polarization
with $\mathrm{F}=S,P$, where $\hat{\mathbf{e}}^S=\hat{\mathbf{s}}$ and
$\hat{\mathbf{e}}^P=\hat{\mathbf{P}}_{v+}$. Replacing
$\mathbf{E}_{\ell}(\omega)\to E_0\mathbf{e}^{\omega}_\ell$, in Eq.
\eqref{eq:tres}, we obtain that
\begin{equation*}\label{eq:m4}
\boldsymbol{\mathcal{P}}_{\ell}(2\omega) = 
\epsilon_{0}E^{2}_{0}\,
\boldsymbol{\chi}_{\mathrm{surface}}:\mathbf{e}^{\omega}_{\ell}
                                     \mathbf{e}^{\omega}_{\ell},
\end{equation*}
where $\mathbf{e}^{\omega}_{\ell}$ is given by Eq. \eqref{eq:newpol},
and thus Eq. \eqref{eq:r10} reduces to ($W_{v}=\cos\theta_{0}$)
\begin{equation*}\label{eq:mr10}
E_{\ell}(2\omega) 
= \frac{i \omega}{c\cos\theta_{0}}
\mathbf{e}^{2\omega,\mathrm{F}}_{\ell}\cdot
\boldsymbol{\chi}_{\mathrm{surface}}:\mathbf{e}^{\omega}_{\ell}
                                     \mathbf{e}^{\omega}_{\ell},
\end{equation*}
in MKS units. For ease of notation, we define
\begin{equation}\label{eq:mc0}
\Upsilon_{\mathrm{F}}(\alpha,\tau)
\equiv 
\mathbf{e}^{2\omega,\mathrm{F}}_{\ell}\cdot
\boldsymbol{\chi}_{\mathrm{surface}}:\mathbf{e}^{\omega}_{\ell}
                                     \mathbf{e}^{\omega}_{\ell},
\end{equation}
where F stands for the outgoing polarization of the SH electric field given by
$\hat{\mathbf{e}}^{\mathrm{F}}$ in Eq. \eqref{eq:r12mm}, and the
$\mathbf{e}^{\omega}_{\ell}\mathbf{e}^{\omega}_{\ell}$ term defines the incoming
polarization of the fundamental electric field as established in Eq.
\eqref{eq:nonmatrix}.

From Eqs. \eqref{eq:rintensities} and \eqref{eq:intensity} we obtain that,
\begin{equation*}\label{eq:r01m}
\begin{split}
\vert E(2\omega)\vert^{2} &=
\vert E_{0}\vert^{4}\frac{\omega^{2}}{c^{2}W^{2}_{v}}\\
\frac{2\epsilon_{0}c|\sqrt{N_{v}}E(2\omega)|^{2}}
     {\left(2\epsilon_{0}c\vert\sqrt{n_{\ell}}E_{0}\vert^{2}\right)^{2}}
&=
\frac{1}{4\epsilon^{2}_0c^{2}}
\frac{2\epsilon_{0}\omega^{2}}{c\cos^{2}\theta_{0}}
\left\vert\frac{\sqrt{N_{v}}}{n^{2}_{\ell}}
\Upsilon_{\mathrm{F}}(\alpha,\tau)\right\vert^{2}\\
\frac{I(2\omega)}{I^{2}(\omega)} &=
\frac{\omega^{2}}{2\epsilon_{0}c^3\cos^{2}\theta_{0}}
\left\vert\frac{\sqrt{N_{v}}}{n^{2}_{\ell}}
\Upsilon_{\mathrm{F}}(\alpha,\tau)\right\vert^{2}\\
\mathcal{R}_{\mathrm{F}}(\alpha,\tau) &=
\frac{\omega^{2}}{2\epsilon_{0}c^3\cos^{2}\theta_{0}}
\left\vert  \frac{1}{n_{\ell}}\Upsilon_{\mathrm{F}}(\alpha,\tau)\right\vert^{2}.
\end{split}
\end{equation*}
Finally, we condense these results and establish the SSHG yield as
\begin{equation}\label{eq:mc6}
\mathcal{R}_{\mathrm{F}}(\alpha,\tau)=
\frac{\omega^{2}}{2\epsilon_{0}c^3\cos^{2}\theta_{0}}
\left\vert\frac{1}{n_{\ell}}\Upsilon_{\mathrm{F}}(\alpha,\tau)\right\vert^{2} 
,
\end{equation}
where $N_{v}=1$ and $W_{v}=\cos\theta_{0}$.
$\boldsymbol{\chi}_{\mathrm{surface}}$ is given in m$^{2}$/V in the MKS unit
system, since it is a surface second order nonlinear susceptibility, and
$\mathcal{R}$ is given in m$^2$/W. 

We now have everything we need to derive the explicit expressions for
$\mathcal{R}$, by using Eqs. \eqref{eq:mc6} and \eqref{eq:mc0}, for any
polarization combination of incoming and outgoing fields. The crux of the matter
now becomes how to calculate Eq. \eqref{eq:mc0}; fortunately, this term can be
expressed in a highly elegant and flexible manner that greatly simplifies the
required algebra. Remember, the four most common combinations of linear
polarizations ($p$-in/$P$-out, $p$-in/$S$-out, $s$-in/$P$-out, and
$s$-in/$S$-out) can be easily recovered from this treatment by using the values
for $\alpha$ and $\tau$ listed in Table \ref{tab:polcases}.

As mentioned before, it will be very convenient to switch all expressions over
to their respective matrix representations. We will start by representing
$\boldsymbol{\chi}$ in this manner. Disregarding all symmetry relations, we have
\begin{equation}\label{eq:chicomplete}
\boldsymbol{\chi} =
\begin{pmatrix}
\chi^{xxx}&\chi^{xyy}&\chi^{xzz} &|& \chi^{xyz}&\chi^{xxz}&\chi^{xxy} \\[3pt]
\chi^{yxx}&\chi^{yyy}&\chi^{yzz} &|& \chi^{yyz}&\chi^{yxz}&\chi^{yxy} \\[3pt]
\chi^{zxx}&\chi^{zyy}&\chi^{zzz} &|& \chi^{zyz}&\chi^{zxz}&\chi^{zxy}
\end{pmatrix}
,
\end{equation}
where all 18 independent components are accounted for, recalling that
$\chi^{\mathrm{abc}} = \chi^{\mathrm{acb}}$ for SHG. Notice that the left hand
block contains the components of $\chi^{\mathrm{abc}}$ where $b = c$, and the
right hand block those where $b \neq c$. If, for example, you have a sample that
is rotated with respect to the original crystal axes, the rotated
$\chi^{\mathrm{abc}}$ components will be a combination of different components
from the original system. In Appendix \ref{app:rot}, we derive the expressions
for the rotated components; they can be substituted directly into the equations
that follow in this section.

Concerning the $1\omega$ terms, we can readily express Eq. \eqref{eq:nonmatrix}
as a combination of vectors,
\begin{equation*}\label{eq:ewsquared}
\mathbf{e}^{\omega}_{\ell}
\mathbf{e}^{\omega}_{\ell}
=
\mathbf{c}\,\mathbf{r}^{\omega}(\alpha,\tau),
\end{equation*}
where
\begin{equation*}
\mathbf{c} = 
\big(
\hat{\mathbf{x}}\hat{\mathbf{x}}\hspace{5pt}
\hat{\mathbf{y}}\hat{\mathbf{y}}\hspace{5pt}
\hat{\mathbf{z}}\hat{\mathbf{z}}\hspace{5pt}
\hat{\mathbf{y}}\hat{\mathbf{z}}\hspace{5pt}
\hat{\mathbf{x}}\hat{\mathbf{z}}\hspace{5pt}
\hat{\mathbf{x}}\hat{\mathbf{y}}
\big),
\end{equation*}
and
\begin{equation}\label{eq:r1wmatrix}
\begin{split}
\mathbf{r}^{\omega}&(\alpha,\tau)
=
\sin^{2}\alpha\,
\left(t^{v\ell}_{s}r^{M+}_{s}\right)^{2}
\begin{pmatrix}
\sin^{2}\phi        \\[8pt]
\cos^{2}\phi        \\[8pt]
0                   \\[8pt]
0                   \\[8pt]
0                   \\[8pt]
- 2\sin\phi\cos\phi
\end{pmatrix}\\[10pt]
&+
e^{2i\tau}\cos^{2}\alpha\,
\left(\frac{t^{v\ell}_{p}}{n_{\ell}}\right)^{2}
\begin{pmatrix}
\left(r^{M-}_{p}\right)^{2}w^{2}_{\ell}\cos^{2}\phi       \\[8pt]    
\left(r^{M-}_{p}\right)^{2}w^{2}_{\ell}\sin^{2}\phi       \\[8pt]
\left(r^{M+}_{p}\right)^{2}\sin^{2}\theta_{0}             \\[8pt]
2r^{M+}_{p}r^{M-}_{p}w_{\ell}\sin\theta_{0}\sin\phi       \\[8pt]
2r^{M+}_{p}r^{M-}_{p}w_{\ell}\sin\theta_{0}\cos\phi       \\[8pt]
2\left(r^{M-}_{p}\right)^{2}w^{2}_{\ell}\sin\phi\cos\phi  
\end{pmatrix}\\[10pt]
&+
2e^{i\tau}\sin\alpha\cos\alpha\,
\frac{t^{v\ell}_{p}t^{v\ell}_{s}r^{M+}_{s}}{n_{\ell}}
\begin{pmatrix}
-r^{M-}_{p}w_{\ell}\sin\phi\cos\phi \\[8pt]
r^{M-}_{p}w_{\ell}\sin\phi\cos\phi  \\[8pt]
0                                   \\[8pt]
r^{M+}_{p}\sin\theta_{0}\cos\phi    \\[8pt]
-r^{M+}_{p}\sin\theta_{0}\sin\phi   \\[8pt]
r^{M-}_{p}w_{\ell}\cos 2\phi        
\end{pmatrix}.
\end{split}
\end{equation}

Likewise, we can obtain the $2\omega$ terms from Eqs. \eqref{eq:e2wpmr} and
\eqref{eq:e2wsmr}, as
\begin{equation*}\label{eq:e2wvector}
\mathbf{e}^{2\omega,\mathrm{F}}_{\ell} =
\mathbf{C}\,\mathbf{R}^{2\omega,\mathrm{F}},
\end{equation*}
where
\begin{equation*}
\mathbf{C} = 
\big(
\hat{\mathbf{x}}\hspace{5pt}
\hat{\mathbf{y}}\hspace{5pt}
\hat{\mathbf{z}}
\big),
\end{equation*}
and
\begin{equation}\label{eq:r2wmatrix}
\mathbf{R}^{2\omega,\mathrm{F}} =
\left\{
\begin{array}{ l }
T^{v\ell}_{p}N_{\ell}^{-1}
\begin{pmatrix}
-R^{M-}_{p}W_{\ell}\cos\phi \\[3pt]
-R^{M-}_{p}W_{\ell}\sin\phi \\[3pt]
+R^{M+}_{p}\sin\theta_{0}
\end{pmatrix},\,\text{for F $ = P$},\\\\\\
T_{s}^{v\ell}R^{M+}_{s}
\begin{pmatrix}
-\sin\phi\\
\cos\phi\\
0
\end{pmatrix},\,\text{for F $ = S$}.
\end{array}
\right.
\end{equation}

Finally, we can express $\Upsilon_{\mathrm{F}}(\alpha,\tau)$ (Eq.
\eqref{eq:mc0}) in complete matrix form as follows,
\begin{equation*}\label{eq:bigassmofo}
\Upsilon_{\mathrm{F}}(\alpha,\tau) =
\mathbf{R}^{2\omega,\mathrm{F}}
\circ\boldsymbol{\chi}\cdot
\mathbf{r}^{\omega}(\alpha,\tau),
\end{equation*}
where we use Eqs. \eqref{eq:chicomplete}, \eqref{eq:r1wmatrix}, and
\eqref{eq:r2wmatrix}, and the ``$\circ$'' symbol is the Hadamard (piecewise)
matrix product. Thus, we select the polarization of the incoming fields via
$\alpha$ and $\tau$ in Eq. \eqref{eq:r1wmatrix}, and the output polarization can
be either $P$ or $S$ in Eq. \eqref{eq:r2wmatrix}. The surface symmetries will be
taken into account via $\boldsymbol{\chi}$ in Eq. \eqref{eq:chicomplete}, or we
can neglect them entirely by calculating every $\chi^{\mathrm{abc}}$ component.
The avid reader will want to consult Refs. \cite{andersonPRB16b} and
\cite{andersonthesis} for the complete derivations of the expressions for
different combinations of linear polarization, for three common surface
symmetries.


\section{Conclusions}

In this manuscript, we have developed complete matrix expressions for the SSHG
radiation using the three layer model to describe the radiating system. This new
treatment now considers the most general polarization case for the incoming
fields, elliptical polarization. It also includes all required components of
$\chi^{\mathrm{abc}}$, regardless of symmetry considerations. Thus, these
expressions can be applied to any surface, regardless of symmetry and for any
choice of incoming polarization. This inherent flexibility of the model makes it
an excellent choice for thin films and 2D materials. Details about the software
implementation of the theory developed here can be found in Ref.
\cite{andersonJOSS17}.


\appendix
\section{Considering an arbitrary rotation on
\texorpdfstring{$\boldsymbol{\chi}(-2\omega;\omega,\omega)$}{X(2w)}}
\label{app:rot}

\begin{figure}[t]
\centering
\includegraphics[width=0.9\linewidth]{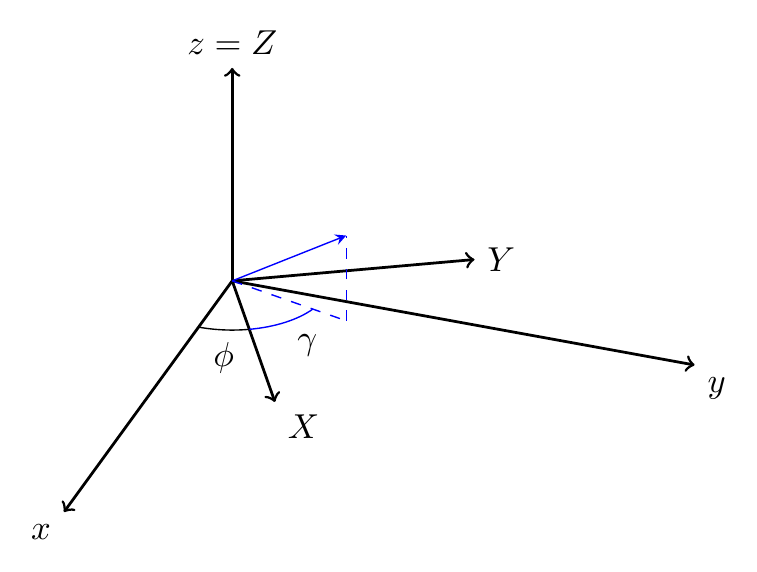}
\caption{The translation from the non-rotated $xyz$ coordinates to the rotated
$XYZ$ system.}
\label{fig:axes}
\end{figure}

To take the components of $\boldsymbol{\chi}(-2\omega;\omega,\omega)$ from the
crystallographic frame to the lab frame, we can simply apply a standard
rotational matrix,
\begin{equation*}
R =
\begin{pmatrix}
R_{Xx} & R_{Xy} & R_{Xz} \\
R_{Yx} & R_{Yy} & R_{Yz} \\
R_{Zx} & R_{Zy} & R_{Zz} \\
\end{pmatrix}
=
\begin{pmatrix}
\sin\gamma & -\cos\gamma & 0 \\
\cos\gamma &  \sin\gamma & 0 \\
    0    &      0    & 1
\end{pmatrix},
\end{equation*}
such that
\begin{equation*}
\chi^{IJK} = \sum_{ijk}R_{Ii}R_{Jj}R_{Kk}\chi^{ijk},
\end{equation*}
where $I$, $J$, and $K$ ($i$, $j$, $k$) cycle through $X$, $Y$, or $Z$ ($x$,
$y$, $z$). Fig. \ref{fig:axes} depicts this rotation over any arbitrary angle
$\gamma$. Since we only consider a rotation in the $xy$-plane along $\gamma$,
the $z$ and $Z$ axes are the same.

Therefore, our $\chi^{IJK}$ components in terms of the original $ijk$ coordinate
system are
\begin{equation*}
\begin{split}
\chi^{XXX} 
&=  \sin^{3}\gamma            \chi^{xxx}\\
&+  \sin\gamma \cos^{2}\gamma \chi^{xyy}\\
&- 2\sin^{2}\gamma \cos\gamma \chi^{xxy}\\
&-  \sin^{2}\gamma \cos\gamma \chi^{yxx}\\
&-  \cos^{3}\gamma            \chi^{yyy}\\
&+ 2\sin\gamma \cos^{2}\gamma \chi^{yxy},
\end{split}
\end{equation*}
\begin{equation*}
\begin{split}
\chi^{XYY} 
&=  \sin\gamma \cos^{2}\gamma \chi^{xxx}\\
&+  \sin^{3}\gamma            \chi^{xyy}\\
&+ 2\sin^{2}\gamma \cos\gamma \chi^{xxy}\\
&-  \cos^{3}\gamma            \chi^{yxx}\\
&-  \sin^{2}\gamma \cos\gamma \chi^{yyy}\\
&- 2\sin\gamma \cos^{2}\gamma \chi^{yxy},
\end{split}
\end{equation*}
\begin{equation*}
\chi^{XZZ} = \sin\gamma \chi^{xzz} - \cos\gamma \chi^{yzz},
\end{equation*}
\begin{equation*}
\begin{split}
\chi^{XYZ} = \chi^{XZY}
&= \sin^{2}\gamma        \chi^{xyz}\\
&+ \sin\gamma \cos\gamma \chi^{xxz}\\
&- \sin\gamma \cos\gamma \chi^{yyz}\\
&- \cos^{2}\gamma        \chi^{yxz},
\end{split}
\end{equation*}
\begin{equation*}
\begin{split}
\chi^{XXZ} = \chi^{XZX}
&=
 - \sin\gamma \cos\gamma \chi^{xyz}\\
&+ \sin^{2}\gamma        \chi^{xxz}\\
&+ \cos^{2}\gamma        \chi^{yyz}\\
&- \sin\gamma \cos\gamma \chi^{yxz},
\end{split}
\end{equation*}
\begin{equation*}
\begin{split}
\chi^{XXY} = \chi^{XYX} 
&= \sin^{2}\gamma \cos\gamma                    \chi^{xxx}\\
&- \sin^{2}\gamma \cos\gamma                    \chi^{xyy}\\
&+ (\sin^{3}\gamma - \sin\gamma \cos^{2}\gamma) \chi^{xxy}\\
&- \sin\gamma \cos^{2}\gamma                    \chi^{yxx}\\
&+ \sin\gamma \cos^{2}\gamma                    \chi^{yyy}\\
&+ (\cos^{3}\gamma - \sin^{2}\gamma \cos\gamma) \chi^{yxy},
\end{split}
\end{equation*}
for the $\chi^{XJK}$ components,
\begin{equation*}
\begin{split}
\chi^{YXX}
&=  \sin^{2}\gamma \cos\gamma \chi^{xxx}\\
&+  \cos^{3}\gamma            \chi^{xyy}\\
&- 2\sin\gamma \cos^{2}\gamma \chi^{xxy}\\
&+  \sin^{3}\gamma            \chi^{yxx}\\
&+  \sin\gamma \cos^{2}\gamma \chi^{yyy}\\
&- 2\sin^{2}\gamma \cos\gamma \chi^{yxy},
\end{split}
\end{equation*}
\begin{equation*}
\begin{split}
\chi^{YYY}
&=  \cos^{3}\gamma            \chi^{xxx}\\
&+  \sin^{2}\gamma \cos\gamma \chi^{xyy}\\
&+ 2\sin\gamma \cos^{2}\gamma \chi^{xxy}\\
&+  \sin\gamma \cos^{2}\gamma \chi^{yxx}\\
&+  \sin^{3}\gamma            \chi^{yyy}\\
&+ 2\sin^{2}\gamma \cos\gamma \chi^{yxy},
\end{split}
\end{equation*}
\begin{equation*}
\chi^{YZZ} = \cos\gamma \chi^{xzz} + \sin\gamma \chi^{yzz},
\end{equation*}
\begin{equation*}
\begin{split}
\chi^{YYZ} = \chi^{YZY}
&= \sin\gamma \cos\gamma \chi^{xyz}\\
&+ \cos^{2}\gamma        \chi^{xxz}\\
&+ \sin^{2}\gamma        \chi^{yyz}\\
&+ \sin\gamma \cos\gamma \chi^{yxz},
\end{split}
\end{equation*}
\begin{equation*}
\begin{split}
\chi^{YXZ} = \chi^{YZX}
&=
 - \cos^{2}\gamma        \chi^{xyz}\\
&+ \sin\gamma \cos\gamma \chi^{xxz}\\
&- \sin\gamma \cos\gamma \chi^{yyz}\\
&+ \sin^{2}\gamma        \chi^{yxz},
\end{split}
\end{equation*}
\begin{equation*}
\begin{split}
\chi^{YXY} = \chi^{YYX}
&= \sin\gamma \cos^{2}\gamma                    \chi^{xxx}\\
&- \sin\gamma \cos^{2}\gamma                    \chi^{xyy}\\
&- (\cos^{3}\gamma - \sin^{2}\gamma \cos\gamma) \chi^{xxy}\\
&+ \sin^{2}\gamma \cos\gamma                    \chi^{yxx}\\
&- \sin^{2}\gamma \cos\gamma                    \chi^{yyy}\\
&+ (\sin^{3}\gamma - \sin\gamma \cos^{2}\gamma) \chi^{yxy},
\end{split}
\end{equation*}
for the $\chi^{YJK}$ components, and lastly
\begin{equation*}
\begin{split}
\chi^{ZXX}
&=  \sin^{2}\gamma        \chi^{zxx}\\
&+  \cos^{2}\gamma        \chi^{zyy}\\
&- 2\sin\gamma \cos\gamma \chi^{zxy},
\end{split}
\end{equation*}
\begin{equation*}
\begin{split}
\chi^{ZYY}
&=  \cos^{2}\gamma        \chi^{zxx}\\
&+  \sin^{2}\gamma        \chi^{zyy}\\
&+ 2\sin\gamma \cos\gamma \chi^{zxy},
\end{split}
\end{equation*}
\begin{equation*}
\chi^{ZZZ} =  \chi^{zzz},
\end{equation*}
\begin{equation*}
\chi^{ZYZ} = \chi^{ZZY} = \sin\gamma \chi^{zyz} + \cos\gamma \chi^{zxz},
\end{equation*}
\begin{equation*}
\chi^{ZXZ} = \chi^{ZZX} = - \cos\gamma \chi^{zyz} + \sin\gamma \chi^{zxz},
\end{equation*}
\begin{equation*}
\begin{split}
\chi^{ZXY} = \chi^{ZYX}
&= \sin\gamma \cos\gamma \chi^{zxx}\\
&- \sin\gamma \cos\gamma \chi^{zyy}\\
&- \cos2\gamma         \chi^{zxy},
\end{split}
\end{equation*}
for the $\chi^{ZJK}$ components. Fortunately, the intrinsic permutation symmetry
of SHG is also present in the new coordinate system, such that $\chi^{IJK} =
\chi^{IKJ}$; therefore, there are only 18 unique components in either system.
Setting $\gamma = \pi/2$ signifies that there is no rotation, and thus
$\chi^{IJK} = \chi^{ijk}$.

\begin{figure*}[t]
\centering
\includegraphics[width=0.5\linewidth]{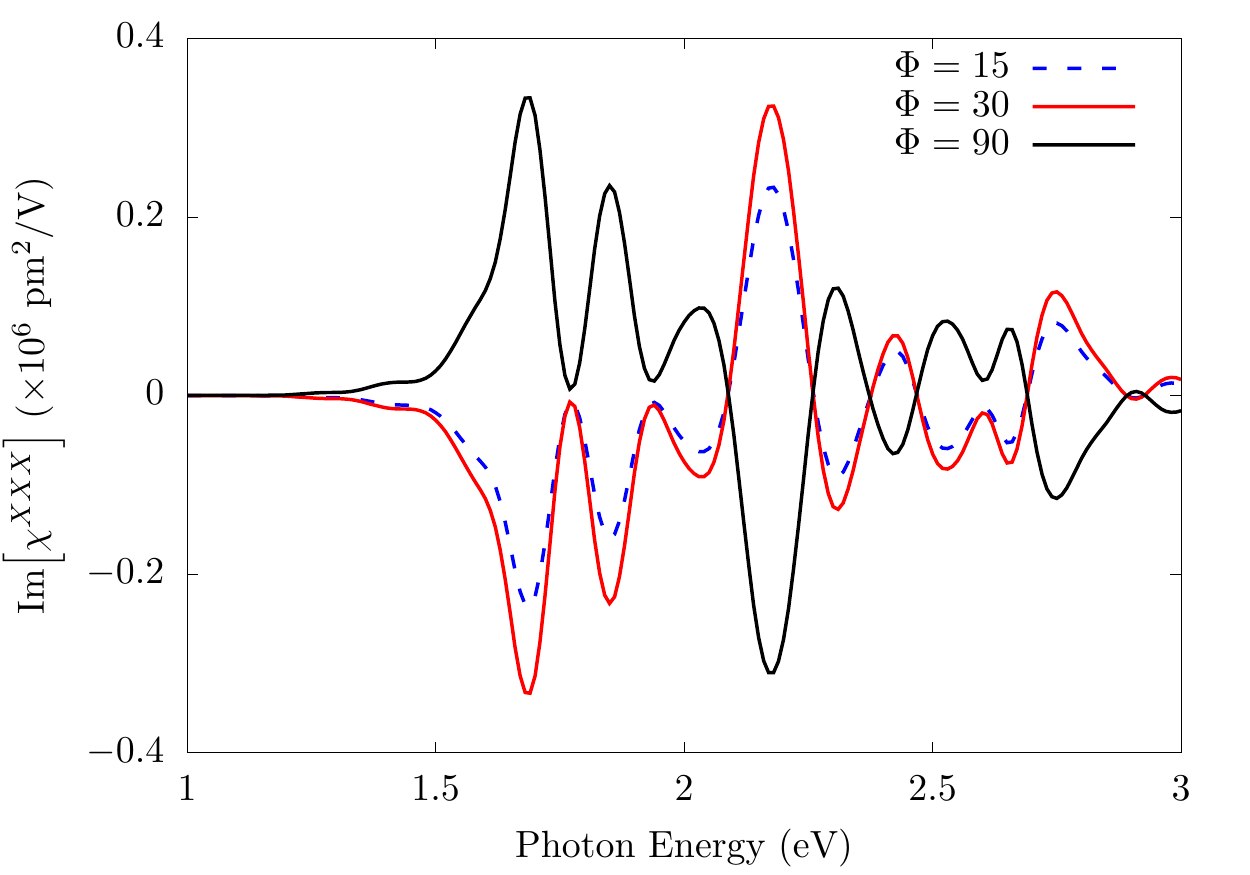}
\caption{$\chi^{XXX}$ for three values of $\gamma$ calculated for a system with
$C_{3v}$ symmetry.}
\label{fig:rotxxx}
\end{figure*}

It should also be clear that the crystal symmetries do \textbf{not} follow into
the rotated system. For instance, the $C_{3v}$ symmetry satisfies the following,
\begin{equation*}
\begin{split}
\chi^{xxx} &= -\chi^{xyy} = - \chi^{yxy},\\
\chi^{yxx} &= \chi^{yyy} = 0.
\end{split}
\end{equation*}
In the rotated system, the top relationship holds true such that $\chi^{XXX} =
-\chi^{XYY} = - \chi^{YXY}$. However, we also obtain that
\begin{equation*}
\chi^{YYY} = \cos3\gamma \chi^{xxx},
\end{equation*}
which is not necessarily zero. Fortunately, we can simply apply the crystal
symmetry to the non-rotated system before transforming to the rotated system. As
an example case, we present $\chi^{XXX}$ for three values of $\gamma$ for a
system with $C_{3v}$ symmetry in Fig. \ref{fig:rotxxx}.
The component in the original coordinates is recoverd when $\gamma = \pi/2$.

\bibliographystyle{unsrt}

\end{document}